\documentstyle[12pt,aasms4]{article}

\slugcomment{To appear in the {\it Astronomical Journal}}

\begin{document}

\title{NGC 5084: A Massive Disk Galaxy Accreting its Satellites ?}

\author{Claude Carignan}
\affil{D\'epartement de physique and Observatoire du Mont M\'egantic,
       Universit\'e de Montr\'eal} 
\authoraddr{C.P. 6128, Succ. `A', Montr\'eal, Qu\'ebec, Canada. H3C 3J7}

\author{St\'ephanie C\^ot\'e}
\affil{European Southern Observatory}
\authoraddr{Karl-Schwarzschild-Str. 2, Garching bei M\"unchen, D-85748,
Germany}

\author{Kenneth C. Freeman}
\affil{Mount Stromlo and Siding Spring Observatories}
\authoraddr{Private Bag PO, Weston Creek, ACT 2611, Australia}

\author{Peter J. Quinn}
\affil{European Southern Observatory}
\authoraddr{Karl-Schwarzschild-Str. 2, Garching bei M\"unchen, D-85748,
Germany}

\begin{abstract}

The spectra of 34 galaxies within 20 arcmins ($\sim$ 100 kpc) of the
lenticular galaxy NGC 5084 have been obtained using the FOCAP
system on the Anglo--Australian 3.9m telescope.
Nine objects are found with projected separations $\lesssim$
80 kpc and with radial velocities within $\pm $ 630 
km s$^{-1}$ of the parent galaxy redshift. 
Using various techniques,
their velocity differences and projected
separations are used to estimate the mass of this S0 galaxy, 
which ranges from 
$6 \times 10^{12} ~M_{\odot}$ to $1 \times 10^{13} ~M_{\odot}$.
With such a mass,
NGC 5084 is one of the most massive disk galaxy known, with a
$(M/L_B) \gtrsim 200\ M_{\odot}/L_{\odot}$. 
In agreement with the models' predictions of Quinn \& Goodman (1986)
but contrary to the results
of Zaritsky {\it et al} (1993) obtained from their statistical sample, the
properties of the satellites' population show no evidence for the
``Holmberg effect'' and a clear excess of satellites in retrograde orbits.
Several signs hint that this S0 galaxy has survived the accretion of several 
satellites. 

\end{abstract}

\section{Introduction}

The nature and distribution of dark matter are still the
most exciting questions of extragalactic astronomy. While
for individual galaxies the mass-to-light ratio $M/L_B$ ranges from 
2 to 90 $M_\odot /L_\odot $ (as derived from their
rotation curves), this ratio shows a wide variety of sometimes
contradictory values when it comes to binary galaxies and small
groups. The most often cited values
of $M/L_B$ for these systems range roughly from 30
up to 600 $M_\odot /L_\odot $ (Huchra and Geller 1982). 
However these values are uncertain 
since many estimates are suspected to be either
polluted by interlopers wrongly assigned to the group, or to
suffer from too stringent selection criteria.
 
A special type of group of galaxies is the one in which a
central dominating massive galaxy is orbited by several dwarf
satellite galaxies. On
inspection of the SERC survey plate, NGC 5084 seems at first
glance to be surrounded by at least two dozens satellites
within 50 kpc (the adopted distance for NGC 5084 throughout
this study is 15.5 Mpc, Zeilinger {\it et al} 1990). So by obtaining the 
redshifts of these objects 
it should be possible to find out the true companions and get rid of most
interlopers. The relative radial velocities and projected separations
of these satellites will then allow an estimate of the mass
of NGC~5084, probing its potential further out than what is possible with an HI
 rotation curve.

Only a handful of studies have used this approach to calculate the mass of
spirals. The difficulty resides in finding enough satellites to carry out such
an analysis. The most populous system studied to date is the group of 5 dwarfs
around NGC~1961 from Erickson {\it et al} (1987). They studied nine groups of 
satellites around spirals (between 1 to 5 satellites each), and concluded 
that only 4 primaries needed a massive dark halo to bind the satellites.
Zaritsky {\it et al} (1993), hereafter ZSFW, conducted a large 
survey of satellites around Sb-Sc's 
and found 69 objects orbiting 45 primaries. Again their most populous system
is NGC~1961 with its 5 dwarfs. But by analysing their whole sample as a 
statistical ensemble they derive a mass of the order 
of $\simeq 2\times 10^{12} ~M_\odot $
 within 200 kpc for a typical Sb-Sc galaxy
similar to the Milky Way (Zaritsky \& White 1994, hereafter ZW),
in good agreement with recent estimates of the mass of the
Milky Way (see for example Freeman, 1996).

This sample of satellites will not only be useful to estimate the
mass of NGC 5084, but it should also be a good test for the
theory of accretion of satellites by a parent disk galaxy elaborated
by Quinn \& Goodman (1986), hereafter QG, who made very definite
predictions on the properties of the left--over satellites' population
after a Hubble time.

NGC~5084 is a lenticular galaxy with a very large
rotational velocity ($\simeq 330$ km s$^{-1}$) and
an extended low surface--brightness disk,
which is tilted at $\sim 5^{\rm o}$ with respect to the brighter,
inner stellar disk. The faint disk extends for at least 6.8
arcmin on both sides of the nucleus. In its inner one-third 
lies a dust lane which passes slightly to the south of the
nucleus. Zeilinger {\it et al} (1990) showed that the
change in position angle describing the tilt between the 2
structures (faint extended disk and bright inner one) is 
present in R as well as in B images, which hints that this
misalignment is intrinsic and not due to dust absorption.
The distorted nature of the inner regions suggests 
the possibility that this galaxy
may be recovering from a recent merger event, probably with what was   
once a dwarf companion. 

Neutral hydrogen was detected in large quantity in NGC~5084.
The study by Gottesman and Hawarden (1986), hereafter
GH, shows that the
HI is distributed in a flat annulus, extending over most of the
faint clumpy disk. The total HI 
mass yields a $M_H/L_B\sim $
0.35 which is very high for a galaxy of this type 
(van Driel \& van Woerden 1991). 
Such a high $(M_H/L_B)$ suggests again the
possibility of a recent dwarf merger(s), with some gas being freshly 
accreted during the event.
Without any further analysis, it is also clear 
from the HI data that NGC~5084 is a
massive galaxy since its derived maximum velocity of 328 km s$^{-1}$ (GH)
is nearly twice the maximum velocity derived for a typical
S0 galaxy (van Woerden, van Driel \& Schwarz 1983).

\section{Observations}

The first step was to identify possible satellites 
by close inspection of the UK Schmidt III--a J survey plate.
Within 22 arcmin of the galaxy (which corresponds to $\sim$100 kpc
at the distance of NGC~5084), 49 candidate satellite galaxies
were identified. The coordinates of these objects were obtained
by using the Space Telescope Guide Star Catalog software and also
by measurements on the survey plate J576 with the Mount Stromlo
PDS microdensitometer.
The positions of the galaxies
were determined to an accuracy of better than 1.2 arcsec rms.
 
With so many spectra to be obtained within such a small region
of the sky, these observations were very well-suited for
the Fibre-Optic Coupled
Aperture Plate system (FOCAP, Boyle {\it et al} 1989)
on the Anglo--Australian Telescope (AAT),
used in conjunction with the Royal Greenwich
Observatory (RGO) Spectrograph and the Image Photon Counting
System (IPCS, Boksenberg 1978).
The observations were carried out on the perfectly clear night of
1990 April 30th. 

{}From the total of 56 fibres available in the bundle, 34 were positionned
on candidate satellites, the remaining 15 satellites 
identified on the plate being either 
vignetted by the TV mirror, either too close from 
a prime candidate, or 
inaccessible within the 40 arcmin field. So the remaining 22 fibres 
were plugged on sky positions. 
The spectra were obtained using the 600
lines/mm grating, with a spectral resolution of 2 \AA ~and a 
spectral coverage of 2000 \AA, in the
range of $\sim$ 3550 to 5550 \AA ~targetting 
the CaII H \& K, H$\beta $, [OIII] and Mg b 
triplet lines. 
Exposures on the program objects of 2000s were alternated with 200s
exposures of a Cu-Ar arc lamp for wavelength calibration, 
for a total of 28 900s on our objects through the night.

\section{Data Reduction}

The data reduction was performed with the software package FIGARO. 
The technique used to obtain the redshifts of the 34 objects was to
cross-correlate their spectra with the ones of 2 template stars,
following the method of Tonry \& Davis (1979). For a handful of objects,
including G8, emission lines were identified in the spectra and 
velocities were obtained by fitting a gaussian to the lines.

Redshifts were succesfully obtained for 28 objects, amongst which 
8 were found to have velocities 
within $\pm 630$ km s$^{-1}$ of NGC 5084 
($V_\odot $ = 1721 km s$^{-1}$), and are thus thought to 
be satellites. These are listed in Table~\ref{t:t1}, with their respective 
projected separation $R$ and relative velocity $\Delta v$ from NGC~5084,
with errors in velocity given by:

\begin{equation}
\sigma _v = \frac{N \times vbin}{8B (1 + R)}
\end{equation}

\noindent where $N$ is the number of channels used in the spectrum, $vbin$ is
the velocity increment per bin, $B$ is the highest wavenumber
where the Fourier Transform of the cross--correlation
function (CCF) has appreciable
amplitude, and $R$ is the ratio of the height of the correlation
peak to the height of an average noise peak in the CCF (Tonry \& Davis 1979). 

ESO576-G40 (G7 in the Table~\ref{t:t1}) is added to our sample of satellites:
although not detected in our FOCAP run as it  
was inadvertently plugged
in a not-known-to-be-dead fibre, it was previously detected in HI by GH, 
at $V_\odot =
2089$ km s$^{-1}$. 
It will be assumed that all these objects
are bound
satellites of NGC 5084. This is a very reasonable assumption
considering that they all have projected separations of less than 
$\simeq $ 80 kpc of NGC~5084 
which, in comparison, has a maximum optical diameter of 74
kpc; and they all have radial velocities within $\pm $ 630 km
s$^{-1}$, which is less than twice the maximum rotational velocity of NGC~5084
(V$_{max} = 328$ km s$^{-1}$). This projected separation cutoff was simply 
dictated by the field-of-view available with FOCAP. As for the relative 
velocity cutoff, it was determined from the observation that the next nearest
$\Delta v$ is 1240 km s$^{-1}$, ie: about twice the value of the last one
retained (G4 with $\Delta v$= 629 km s$^{-1}$). It therefore seems to indicate
that NGC~5084 and its satellites' system have decoupled from the general 
Hubble flow and can be safely considered a self-gravitating ensemble. 

The satellites are identified in Figure 1 and the radial velocity
differences between the satellites and NGC~5084 are shown in Figure 2
as a function of their projected separation. Figure 3, which gives
the $\Delta v$'s of all the candidates (satellites and background
galaxies), shows clearly that this satellites' system is well
isolated. While the retained satellites all have $\Delta v < 630$
km s$^{-1}$, most of the other observed galaxies have $\Delta v > 10 000$
km s$^{-1}$ with only 3 objects with $1200 < \Delta v < 5300$ km s$^{-1}$.

\section{Mass estimate of NGC 5084}

The radial velocities derived in the last section will now be used
to estimate the mass of NGC 5084.
From its maximum velocity
of $V_{max}$ = 328 km s$^{-1}$ and the outermost measured radius
of 7.5 arcmin (34 kpc), 
a Keplerian mass estimate (~$rV^2/G$~) 
of $M = 8.5 \times
10^{11} ~M_\odot $ is derived, leading to an exceptionnally large
mass-to-light ratio of $M/L_B = 65\ M_\odot /L_\odot $.
NGC~5084 is clearly one of these supermassive disk galaxies,
as referred to by Saglia and Sancisi (1988). Several galaxies that they 
list have even higher $V_{max}$, but NGC~5084 is the first one for which
dwarf satellites are going to be used to probe the potential further out.

Let us first apply the virial theorem to estimate the mass of NGC
5084, which is the standard method for obtaining the mass of a
self-gravitating system. We will be using 8 of the 9 objects listed 
in Table~\ref{t:t1}, discarding G8 which 
appears to be a satellite of a satellite! 
It is closely orbiting G7 and so 
its kinematics surely reflects more its interaction with G7 than with 
NGC~5084.
In the special case of a spherically symmetric
collection of N test particles orbiting a point mass M, the virial
theorem takes the form of (see Bahcall and Tremaine 1981, hereafter BT, for 
the derivation):

\begin{equation}
M_{VT} = {3 \pi \over 2 G} (\sum_{i=1}^N \Delta v_i^2)\bigg/ (\sum_{i=1}^N
1/R_i)
\end{equation}

\noindent Using the values of $\Delta v_i$ and $R_i$  (and their errors) 
of Table~\ref{t:t1}, the virial mass is:

$$M_{VT} = 6.3 (\pm 3.4) \times 10^{12}  ~M_\odot $$

However, BT argue that this virial estimator is
biased and inefficient. They show that $<M_{VT}>$ is not necessarily
equal to M for finite $N$ and that $M_{VT}$ does not converge to M as
$N\to \infty $. They also complain  that it weights the contribution from nearby
particles too heavily. They have therefore proposed alternative
estimators based on the projected mass, q = (projected distance) (radial
velocity) /G. These estimators take the following forms:

\begin{equation}
M_{BT} = {f\over \pi G N} \sum_{i=1}^N  \Delta v_i^2 R_i 
\end{equation}

\noindent where $f$ = 16 in the case of isotropic orbits, and 
$f$ = 32 for radial orbits. 
Since there is no information on the
distribution of eccentricities for the satellites' orbits, it is
recommended to use the following estimator:

\begin{equation}
M_0 = 1/2 ~(M_{iso} + M_{rad}) = {24\over \pi G N} \sum_{i=1}^N  \Delta v_i^2 R_i
\end{equation}

\noindent where $M_{iso}$ is $M_{BT}$ for the isotropic case and $M_{rad}$ for 
the radial case.  

Applying the BT estimators to our sample, these yield:

$$M_{iso} = 6.7\ (\pm 3.5) \times 10^{12}  ~M_\odot $$
$$M_{rad} = 1.3\ (\pm 0.7) \times 10^{13}  ~M_\odot $$
$$M_0 = 1.0\ (\pm 0.5) \times 10^{13}  ~M_\odot $$

\noindent BT have also carried out a series of
Monte Carlo simulations from which they concluded that the
virial mass is too often an underestimate, and that their
projected mass estimator is more accurate. 
This BT estimator and the virial one share however some limitations. 
First, these estimators assume random phases, ie: the orbital
phases of the satellites are taken to be uniformly and 
independently distributed on
[0,2$\pi $]. This is a good approximation only 
if the satellites have completed many
orbits. Indeed calculating the period of a test 
particle at 82.95 kpc (our maximum 
projected separation) with a velocity 
of 328 km s$^{-1}$ (=V$_{max}$ of NGC 5084) gives about
1.5$\times $10$^9$ years, allowing for several 
orbits within a Hubble time, and so 
this random-phases assumption is appropriate for our sample.
Second the potential is assumed to be generated by 
a central point-mass, which is inadequate
here as the satellites are at very small projected separation and are probably 
orbiting within the dark halo of NGC 5084.
ZW circumvent this limitation by applying to their 
statistical sample of satellites a more sophisticated method using scale-free
models, originally developped for dynamical studies of binary galaxies (Turner 
1976; White 1981). The underlying assumption 
of these models  is that small-scale
galaxy clustering exhibits no characteristic length-scale. The interaction 
potential and the density profile of satellites 
are taken to be power-laws. As will
be discussed further, N-body simulations of accretion of satellites by a parent
disk galaxy make definite predictions on the 
properties of the left--over satellites' population 
after a Hubble time. They find 
very short decay times for prograde orbits satellites (QG).
This is especially valid for satellites at 
distances within a disk-diameter from their 
primary as is roughly the case here. So under these conditions one must perhaps 
question the appropriateness of the scale-free assumption. 
In any case ZW
found good agreement between the BT mass estimates and the scale-free
models estimates. 
Therefore our best estimate
for the mass of NGC 5084 is $\sim 1.0\times 10^{13} ~M_\odot $,
 with an uncertainty of a factor of 2.
{\it This is the highest mass yet ever derived for a disk
galaxy}.

We have recalculated these estimates considering the possibility that the 2
most extreme objects (in terms of $\Delta v^2 R$), G4 and G7, are not bound
to the system. We see from Figure 2 that these two galaxies lie well above
general trend defined by the rest of the sample, so it is worth considering
the consequences of their exclusion. The remaining 6 satellites then give:
$$ M_{VT} = 3.9 \times 10^{12} M_\odot$$  $$M_0 = (5.2 \pm 2.9) \times
10^{12} M_\odot$$ which is still {\it the highest known mass for a disk
galaxy}.
 
Taking for NGC 5084 the luminosity derived by Zeilinger {\it et al} (1990)
of $L_B = 1.6 \times 10^{10}  ~L_\odot $, a very large
mass-to-light ratio of $(M/L_B)$ = 215 $M_\odot /L_\odot $ is derived.
This indicates
a tremendous amount of dark matter associated with this galaxy. 
The increase of more than a factor of 3 of the $(M/L_B)$
over the value derived by GH is not surprising since this
study is probing the mass distribution and the dark halo
component to much greater radii ($\sim 80$ kpc) than the
last measured point of the HI rotation curve ($\sim 34$ kpc).
The main uncertainty on this ratio is probably coming from
the correction for extinction due to dust within the 
galaxy which could have been underestimated. For example,
if the total magnitude was 0.5 mag. brighter, this would reduced
the $(M/L_B)$ to 135. However, the IRAS fluxes measured for
NGC 5084 (Knapp {\it et al.} 1989) suggest a normal dust
content for its morphological type.

\section{Discussion}

\bigskip
As mentioned earlier, Saglia and Sancisi (1988) have compiled
a list of the most supermassive disk galaxies known.
However these masses are only
indicative masses ($M= R_{25} \times V_{max}^2 /G$) and it is more than
probable that if a similar analysis as the one done here for NGC 5084 could
be carried out, a
few of these would probably be even more massive
then NGC 5084.
 
It is interesting to note that several of these supermassive galaxies
are reported to be clearly asymmetric in their light and/or in their HI
distribution. As we have seen, NGC 5084 is a disk galaxy with an
intrinsically distorted structure, and therefore its
anomalies make it typical of its supermassive class. This has led Saglia
and Sancisi (1988) to propose that supermassive spirals might be forming
a distinct category of galaxies with definite properties instead of
being viewed as the extreme tail of the mass distribution.
These galaxies have most probably accreted some of their satellites, which have 
during the process disturbed the stellar disk. 
Recently, Huang \& Carlberg (1995) have
indeed shown with self-consistent disk+halo+satellite 
N-body simulations that the 
disks of the primaries get mainly tilted (rather than thickened) by infalling 
satellites. This is because large satellite orbital angular momenta which are 
not aligned with disk rotational angular momenta can easily tilt the disks, and
tidal stripping of satellites in the dark halos can 
greatly weaken the satellite's
impact on the disks. A 5$^{\rm o}$ tilt as is 
observed in NGC 5084 would correspond 
to the infall 
of a satellite as massive as roughly 16\% of 
the disk-mass of NGC 5084, according to 
their simulations.

Let us now compare our results with the model's predictions of QG,
and with the results of ZSFW
using their statistical sample. First, the ``Holmberg
effect''. Holmberg (1969), studying a sample of 
218 apparent companions at less than 
$R \leq 50$ kpc 
of 58 parent spiral galaxies, claimed that satellites seem to avoid
a zone of $\pm 30^{\rm o}$ from the plane of the primary which could suggests
that they are the systems which are 
preferentially accreted. ZSFW
also said that their sample shows some evidence for the ``Holmberg
effect''. Looking at Table~\ref{t:t1} and Figure~4, 
there is surely no evidence of
this effect in this restricted sample
with half of the satellites within the $\pm 30^{\rm o}$ zone.
This effect was not seen either in the simulations of QG.
There is no clear indication either in our 
sample of any trends in the satellites
properties as their projected separation increases, 
namely if their sizes increase,
or if they become of lower surface-brightness, or if they tend to be of later 
Hubble types (dIrrs), all of these three trends being 
mildly present in the ZSFW 
sample.

Finally, another prediction of the QG's theory of satellites accretion is
that there should be a net excess of satellites in retrograde orbits
since the parent galaxy should have accreted most of the satellites
in direct orbits within a Hubble time. They found in their N-body simulations 
that the decay times are a factor of 10 times larger for bodies in retrograde 
orbits than for those in direct orbits. Thus, the majority of satellites on 
prograde orbits should have already been accreted within a Hubble time, 
while most of
the ones in retrograde orbits should still 
be at reasonable distances from the primary galaxy.
The final column of Table~\ref{t:t1} compares 
the {\it sign} of $\Delta v$ for the
satellite, and the sign of the rotational velocity of the HI in NGC 5084 on
the same side of the minor axis as the satellite. (P) denotes that the
satellite velocity and the HI velocity have the same sign, while (R) denotes
that the signs are opposite. We cannot infer from an (R) or a (P) whether an
individual satellite orbit is retrograde or prograde.  However it is clear
that 7 of our 8 satellites have an (R) in Table~\ref{t:t1}, 
which does suggest a
predominance of retrograde orbits. This is in contrast to the results of
ZSFW who found nearly equal numbers of (R) and (P) orbits.

\section{Summary and Conclusions}

{}From the spectra of 34 objects identified to be lying close to NGC~5084,
8 galaxies are found with redshifts within $\pm
630$ km s$^{-1}$ of this system, and projected
separations $\leq$ 80 kpc of it. They are used to
probe the potential of NGC 5084. Using 
the projected mass estimator,
a mass of $M \simeq 1.0 \times 10^{13}  ~M_\odot $ is calculated,
therefore
making of NGC 5084 the most supermassive disk galaxy known so far. Such a huge
mass strongly suggests the existence of a considerable quantity
of dark matter associated with NGC 5084. 
Zeilinger {\it et al} (1990) already proposed
that NGC 5084's massive structure might have been built by
sequential accretions of mass from its environment. With so many
close dwarf companions this possibility cannot be excluded.

An analysis of the properties of the satellites' sample shows
no sign of the ``Holmberg effect'' and a clear preference for
retrograde orbits. This agrees with QG's N-body simulations who find that 
parent galaxies do not accrete preferably 
satellites with $\Theta \leq \pm 30\deg $ from the plane of their disks, but
accrete more efficiently those with direct (prograde) orbits.
The fact that most of the prograde orbits satellites have disappeared, that 
NGC~5084 is unusually HI--rich, and that its outer disk is tilted (which is the
signature of small-mass satellites 
accretion according to Huang \& Carlberg 1995 
N-body simulations), all hint strongly that this S0 has survived succesfully 
the accretion of many low-mass objects.

\acknowledgments

We are grateful to Stephen Gottesman for his referee's report,
which improved the content of this paper.
We would like to thank Sylvie Beaulieu and the Space Telescope's
Guide Star Catalog group for providing the software and the assistance
in determining the objects' accurate positions. S.C. acknowledges the
support of an A.N.U. scholarship as well as a F.C.A.R. scholarship, 
and C.C. a grant from NSERC.

\clearpage

\begin{deluxetable}{lcclcc}
\tablecaption{Orientation parameters of the satellites.\label{t:t1}} 
\tablehead{
\colhead{Object} 	  &
\colhead{R.A.~~~Dec}      &
\colhead{Projected}       &
\colhead{Relative}        &
\colhead{Angle from}      &
\colhead{Prograde or}      \\
\colhead{}                &
\colhead{(1950.0)}        &
\colhead{separation}      &
\colhead{radial velocity} &
\colhead{the plane}       &
\colhead{Retrograde}       \\
\colhead{}                &
\colhead{}                &
\colhead{(kpc)}	          &
\colhead{(km s$^{-1}$)}	  &
\colhead{($^{\rm o}$)}    &
\colhead{orbits}
} 

\startdata

G1& 13 16 40.1 -21 20 27  &82.95	&$-$34 ($\pm 96$)	&60	&R   \nl
G2& 13 16 47.4 -21 31 42  &50.15	&$-$232 ($\pm 136$)	&25	&R \nl
G3& 13 17 04.4 -21 38 16  &37.48	&$-$260 ($\pm 57$)      &22     &R \nl
G4& 13 17 29.7 -21 44 26  &48.00	&$-$629 ($\pm 123$)	&71	&R   \nl
G5& 13 17 54.5 -21 37 03  &25.70	&+485 ($\pm 118$)	&46	&R   \nl
G6& 13 17 59.7 -21 30 21  &30.94	&$-$381 ($\pm 125$)	&16	&P   \nl
G7& 13 18 00.9 -21 47 21  &66.98	&+368 ($\pm 3$)	&79	&R   \nl
G8\tablenotemark{*}& 13 18 06.7 -21 47 35  &70.65	&+359 ($\pm 6$)	&---	&--- \nl
G9& 13 18 20.4 -21 29 24  &52.29	&+173 ($\pm 90$)	&9	&R   \nl

\tablenotetext{*}{taken out of the sample}
\enddata
\end{deluxetable}

\clearpage
\begin{figure}
\figurenum{1}
\plotone{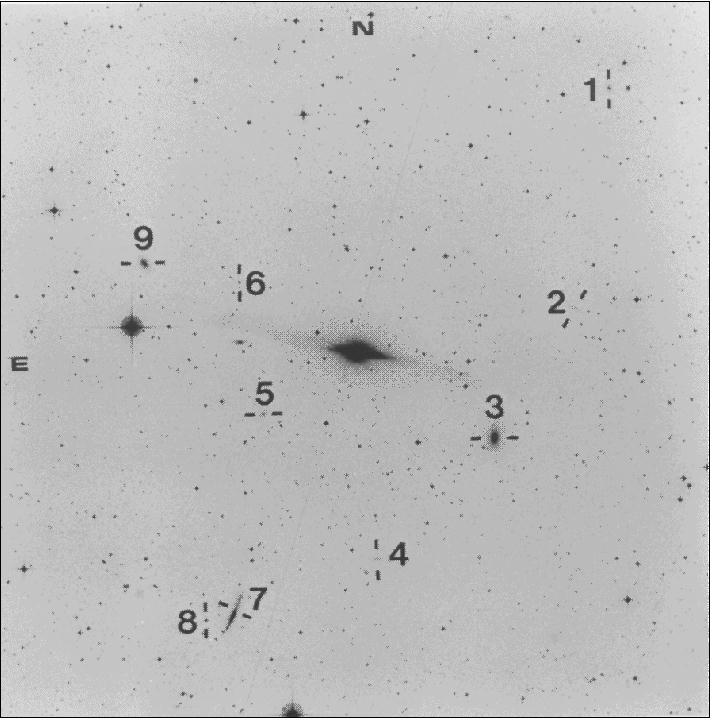}
\caption{Optical identifications for the satellites of Table 1.}
\end{figure}

\clearpage
\begin{figure}
\figurenum{2}
\plotone{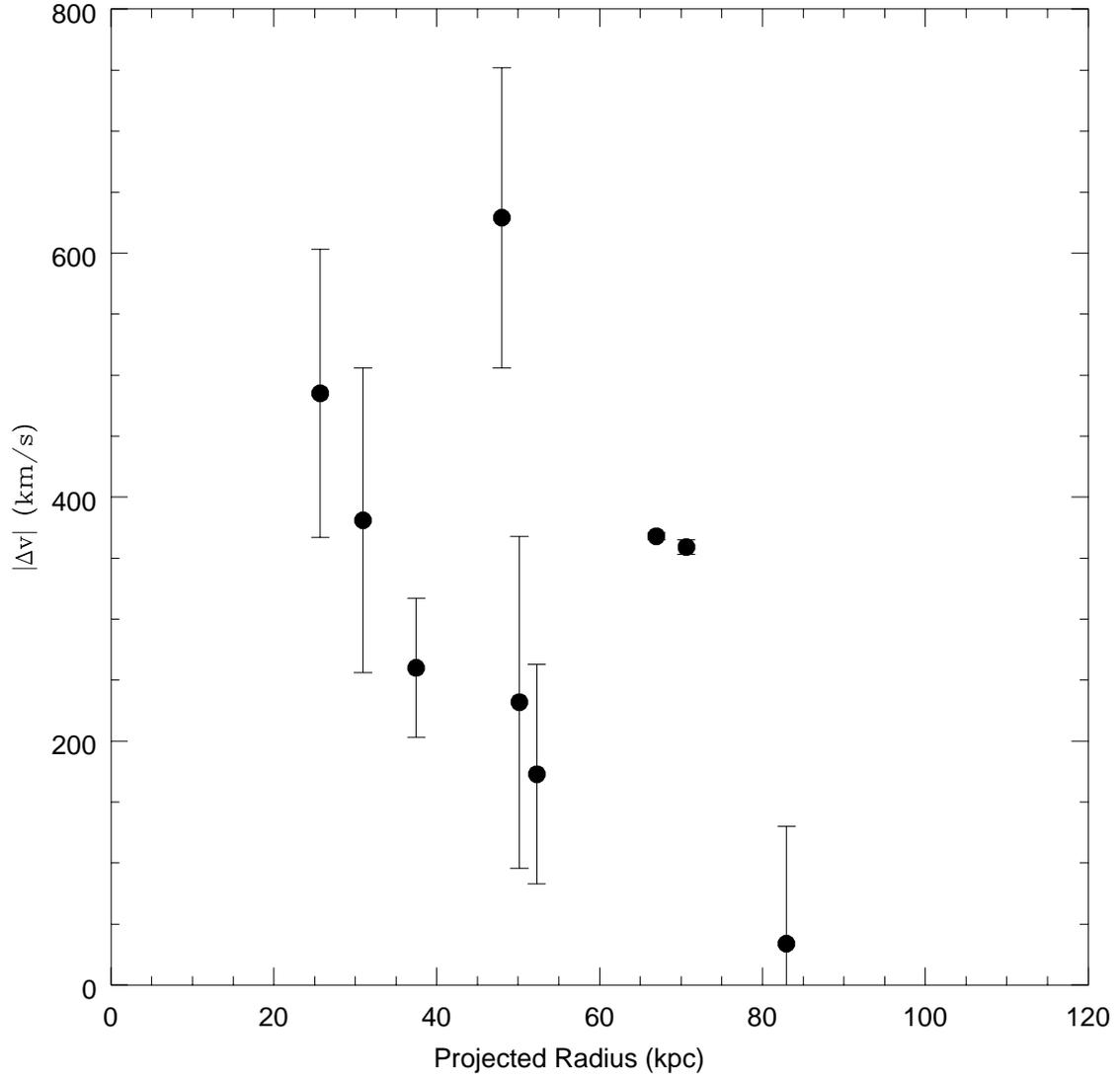}
\caption{The absolute value, $|\Delta v|$, of the radial velocity difference
between a satellite and NGC~5084, as a function of its projected separation
on the sky.}
\end{figure}

\clearpage
\begin{figure}
\figurenum{3}
\plotone{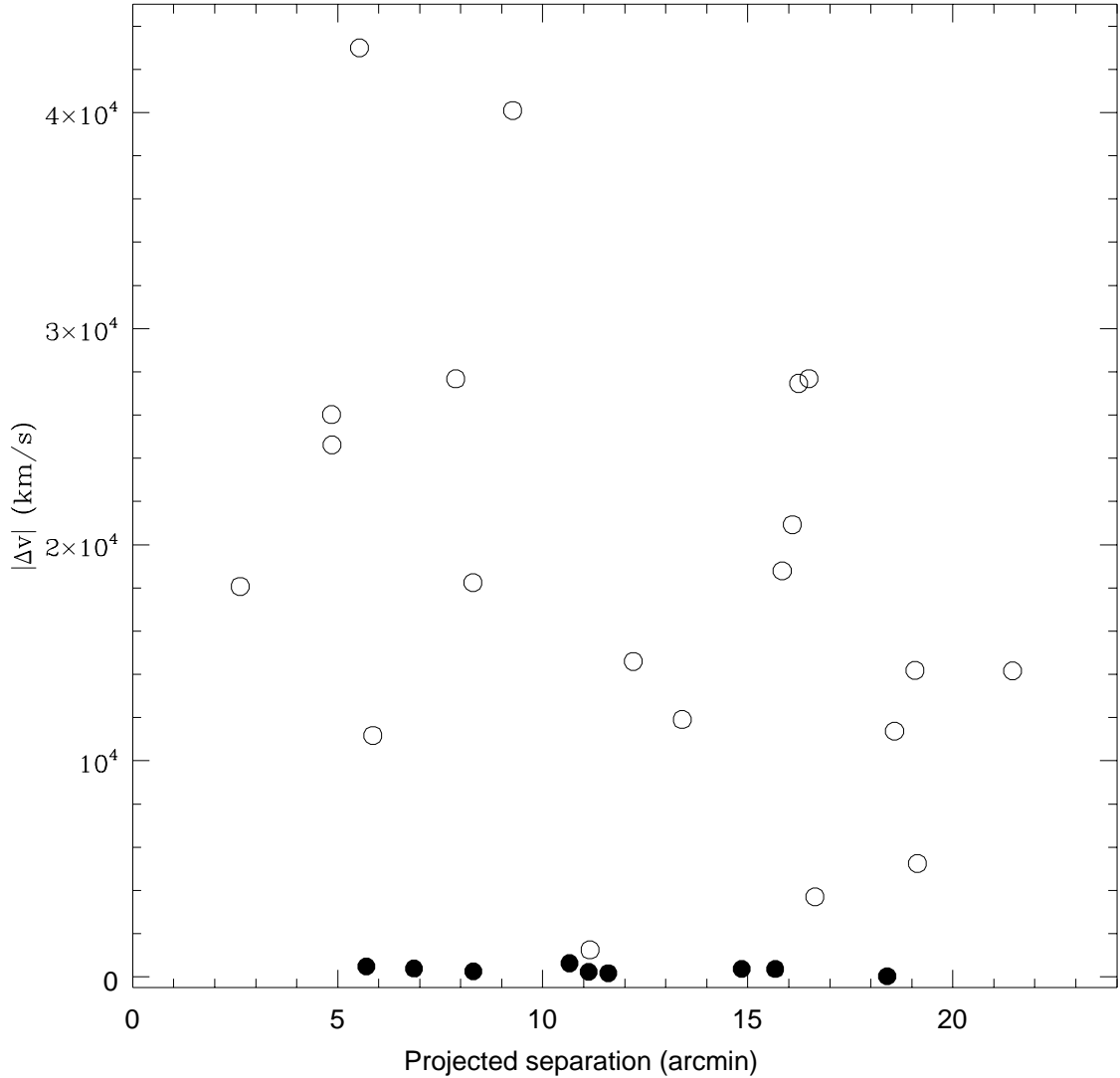}
\caption{$\Delta v$ vs projected separation in arcmin for the 
28 detected objects and for G7 which had a previously known redshift.
The filled circles are the objects identified as satellites
while the open circles are the background galaxies.
The satellites have $\Delta v$'s $< 630$ km s$^{-1}$
while most background galaxies have
$\Delta v >$ 10 000 km s$^{-1}$ with only 3 galaxies 
with redshift between 1200 and
5300 km s$^{-1}$.}
\end{figure}

\clearpage
\begin{figure}
\figurenum{4}
\plotone{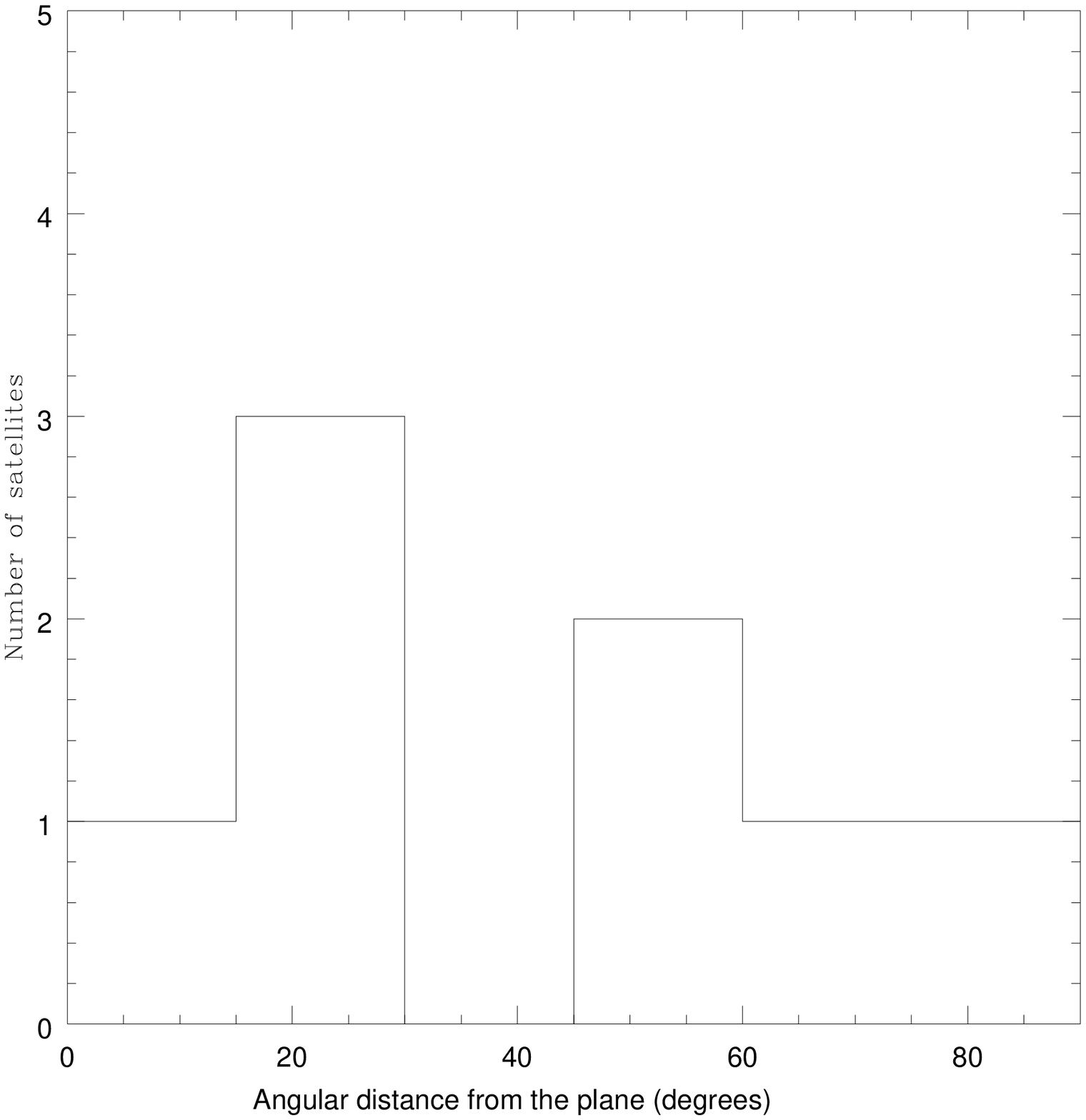}
\caption{Histogram of the number of satellites as a function of the
angular distance from the plane. It can be seen that half of the
satellites are in the $\pm30^{\rm o}$ zone of avoidance predicted by
the ``Holmberg effect''.}
\end{figure}


\begin{references}

\reference {}Bahcall, J.N., \& Tremaine, S. 1981, \apj, 244, 805 (BT)

\reference {}Boksenberg, A. 1978, Report: Performance of the UCL IPCS

\reference {}Boyle, B., Gray, P., \& Sharples, R. 1989, AAO FOCAP System

\reference {}Erickson, L.K., Gottesman, S.T., \& Hunter Jr., J.H. 1987, \nat, 325, 779 

\reference {}Freeman, K.C. 1996, in {\it Unsolved Problems of the Milky
Way}, ed. L. Blitz \& P. Teuben (Dordrecht: Kluwer), p. 645

\reference {}Gottesman, S.T., \& Hawarden, T.G. 1986, \mnras, 219, 759 (GH)

\reference {}Holmberg, E. 1969, Arkiv Astr., 5, 305

\reference {}Huang, S., \& Carlberg, R.G. 1995, Preprint astro-ph/9511076, submitted 
to \apj 

\reference {}Huchra, J., Geller, M., 1982, \apj, 257, 423

\reference {}Knapp, G.R., Guhathakurta, P., Kim, D.--W. and Jura, M. 1989, 
\apjs , 70, 329

\reference {}Quinn, P.J. \& Goodman, J. 1986, \apj, 309, 472 (QG)

\reference {}Saglia, R.P., \& Sancisi, R. 1988, \aap, 203, 28

\reference {}Tonry, J.R., \& Davis, M. 1979, \aj, 84, 1511

\reference {}Turner, E.L. 1976, \apj, 208, 304

\reference {}van Driel, W., \& van Woerden, H., 1991, \aap, 243, 71

\reference {}van Woerden, H., van Driel, W. \& Schwarz, U.H. 1983, in 
{\it IAU Symposium 100, Internal Kinematics and Dynamics of Galaxies}, ed. E. 
Athanassoula (Dordrecht: Reidel), p. 99

\reference {}White, S.D.M. 1981, \mnras, 195, 1037 

\reference {}Zaritsky, D., Smith, R., Frenk, C. \& White, S.D.M. 1993, \apj, 
405, 464 (ZSFW)

\reference {}Zaritsky, D., \& White, S.D.M. 1994, \apj, 435, 599 (ZW)

\reference {}Zeilinger, W.W., Galetta, G., \& Madsen, C. 1990, \mnras, 246, 324

\end{references}
\end{document}